\documentclass{article}

\PassOptionsToPackage{numbers, compress}{natbib}

    \usepackage[preprint]{neurips_2024}

\usepackage{hyperref}
\usepackage{multirow}
\usepackage[table]{xcolor}
\usepackage{hyperref}
\usepackage{graphicx}
\AtBeginDocument{%
  }

\title{Beyond Release: Access Considerations for Generative AI Systems }

\author{
  Irene Solaiman\thanks{First author. All subsequent authors listed in alphabetical order by last name. Inclusion as an author does not entail endorsement of all aspects of the paper. Correspondence to: irene@huggingface.co} \\
  Hugging Face \\
\And Rishi Bommasani \\
  Stanford University \\
\And Dan Hendrycks \\
  Center for AI Safety\\
\And Ariel Herbert-Voss \\
  RunSybil \\
\And Yacine Jernite \\
  Hugging Face \\
\And Aviya Skowron \\
  EleutherAI \\
\And Andrew Trask \\
  OpenMined \\
}

\begin{document}

\maketitle

\begin{abstract}
Generative AI release decisions determine whether system components are made available, but release does not address many other elements that change how users and stakeholders are able to engage with a system. Beyond release, access to system components informs potential risks and benefits. Access refers to practical needs, infrastructurally, technically, and societally, in order to use available components in some way. We deconstruct access along three axes: resourcing, technical usability, and utility. Within each category, a set of variables per system component clarify tradeoffs. For example, resourcing requires access to computing infrastructure to serve model weights. We also compare the accessibility of four high performance language models, two open-weight and two closed-weight, showing similar considerations for all based instead on access variables. Access variables set the foundation for being able to scale or increase access to users; we examine the scale of access and how scale affects ability to manage and intervene on risks. This framework better encompasses the landscape and risk-benefit tradeoffs of system releases to inform system release decisions, research, and policy.

\end{abstract}

\maketitle

\section{Introduction}
Release decisions for generative AI systems raise ongoing discourse, debates, and regulatory questions. Sometimes framed as an “open versus closed” debate, release considerations \cite{solaiman_release_2019} are discussed in media \cite{isaac_what_2024}, a U.S. government executive order \cite{house_executive_2023}, and research on release decisions \cite{brammer_how_2023}. A central issue is determining responsible release that balances tensions of benefits and risks along the gradient of fully open and fully closed systems \cite{solaiman_gradient_2023}. Research and policy have focused on how a model is released, availability of model weights \cite{national_telecommunication_and_information_administration_dual-use_2024}, and more recently, what system components are released \cite{basdevant_towards_2024, liang_time_2022, noauthor_open_2024}, such as model weights, training data, and training code. 

Discourse and analysis has been too narrowly scoped; instead, focus should include aspects of access. Release analysis cannot only consider whether a system and its components are made available, but also must consider how accessible each component is. Release and access to AI systems are often used interchangeably, but access includes the resources and qualities needed for stakeholders to engage with system components. Overall access, beyond release, more concretely determines outcomes. 

Generative AI release decisions for whether a system component is made available and to whom it is available do not provide the full information for long-term risk-benefit considerations. Once components are released, making those components accessible determines whether they can be used and enables scaling usage and reach. For example, due to resourcing, a compute-intensive open-weight model may not be accessible to many researchers. Due to technical usability, user interfaces can make both open- and closed-weight models more accessible to malicious actors with low computer literacy. Due to utility, less comprehensive documentation can limit research. Accessibility makes systems more scalable, affecting who has access and resources needed for risk management. 

In this paper, we explain how analyzing variables beyond release is an effective means of weighing generative AI benefits and risks, in order to inform deployer release decisions, policymaker actions, and further research. We differentiate system release and availability of components from how a system and its components are made accessible. Access goes beyond what is available. The three axes of access shed light on how people and society can be helped or harmed by generative AI systems: resourcing, usability, and utility. Each subsection examines risks and benefits per system component. We examine who can access what component and how scaling access affects managing deployment. 

\section{Previous and Related Work}
Existing work to distinguish between how a system is released and how it interacts with the public includes defining release as making model components available and deployment as the vector of impact \cite{howard_what_2024}. Research to prevent AI misuse explains the misuse-use tradeoff, showing a misuse chain where harm occurs after release \cite{anderljung_protecting_2023}. Laying out the model pipeline gives another approach to considering certain post-release variables \cite{eiras_risks_2024}. Examining openness must also include the resourcing and materials around a system \cite{widder_why_2024}. Governance proposals specific to open foundation models highlight downstream use considerations and the impact of open systems on the overall AI ecosystem \cite{bommasani_considerations_2024}. Recent papers analyze open systems in sectoral contexts, spanning from national security and defense priorities \cite{dahlgren_defense_2024} to fact-checking organizations \cite{wolfe_implications_2024}. Motivations to make systems more open cite scientific reproducibility, product reliability and data security, and economic diversification. Motivations toward closedness per component cite inability to monitor models with downloadable weights and less control to intervene or revoke access to a given component. 

There has been a lack of consensus on key terms such as “open source”, misuse of the term, and popularization of the term “openness” \cite{lambert_why_2023}. Works toward concrete terminology lay out openness dimensions \cite{basdevant_towards_2024}, create an openness framework \cite{white_model_2024}, and propose a formal “open source AI” definition \cite{noauthor_open_2024}. Further work explores  open foundation model marginal risk \cite{kapoor_societal_2024}, safety specific to open foundation models \cite{neumann_prism_2024}, and risk mitigation for open foundation models \cite{srikumar_risk_2024}. 

\section{From Available to Accessed}
\begin{figure}[h!]
  \centering
  \includegraphics[width=\textwidth]{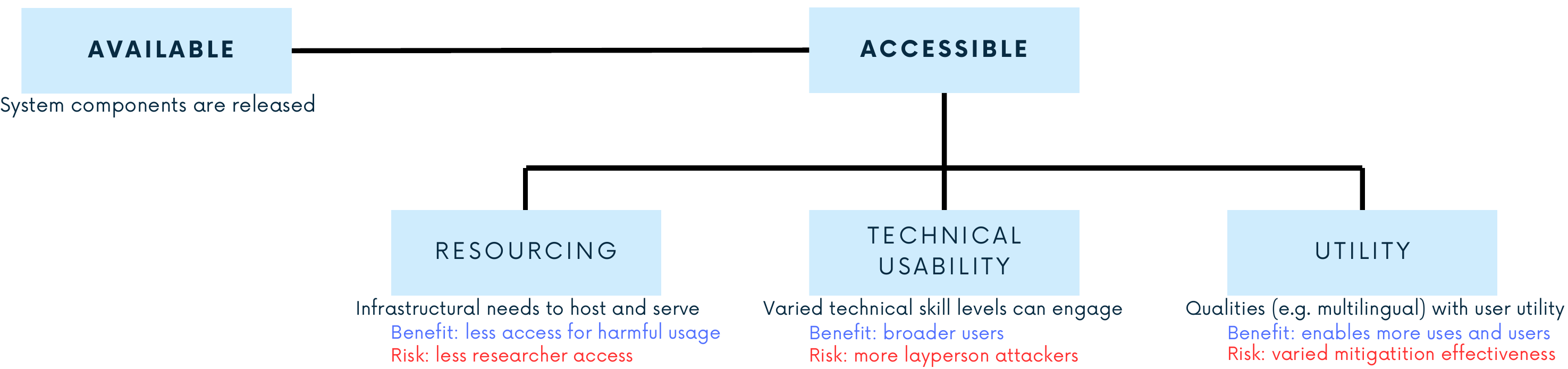}
  \caption{From Available to Accessible with High Level Tradeoffs}
  \label{fig:availaccess}
\end{figure}

Release methods commonly address what is released or  made available. The gradient of release lays out availability: whether or not a system and its components are released along a spectrum \cite{solaiman_gradient_2023}. Whether a system’s components are released or made open is distinct from how accessible the system is; a system may be fully openly released, but not accessible to the many groups affected. When system components are released, they are not useful until they are made accessible. Fig. \ref{fig:availaccess} distinguishes availability and accessibility and shows the three subsections of access. Release can be viewed as the initial conditions for whether a component is available before accounting for making the components accessible to an external stakeholder. Accessible subsections grapple with tradeoffs per component, with high level risks and benefits noted in Figure \ref{fig:availaccess}.

Access variables set the grounding necessary for scaling access; from individuals to a broader audience, ensuring components are accessible enables usage. The scale of distribution of components is a critical factor for better understanding the landscape of accessible system components, both for research and product. Who and how many people are able to access and deploy systems influences scale-related tradeoffs, such as how to manage potential harmful usage. See Appendix \ref{Additional Figures} for additional figures and visual representations.

\section{Breaking Down Access}
\begin{figure}[h!]
  \centering
  \includegraphics[width=\textwidth]{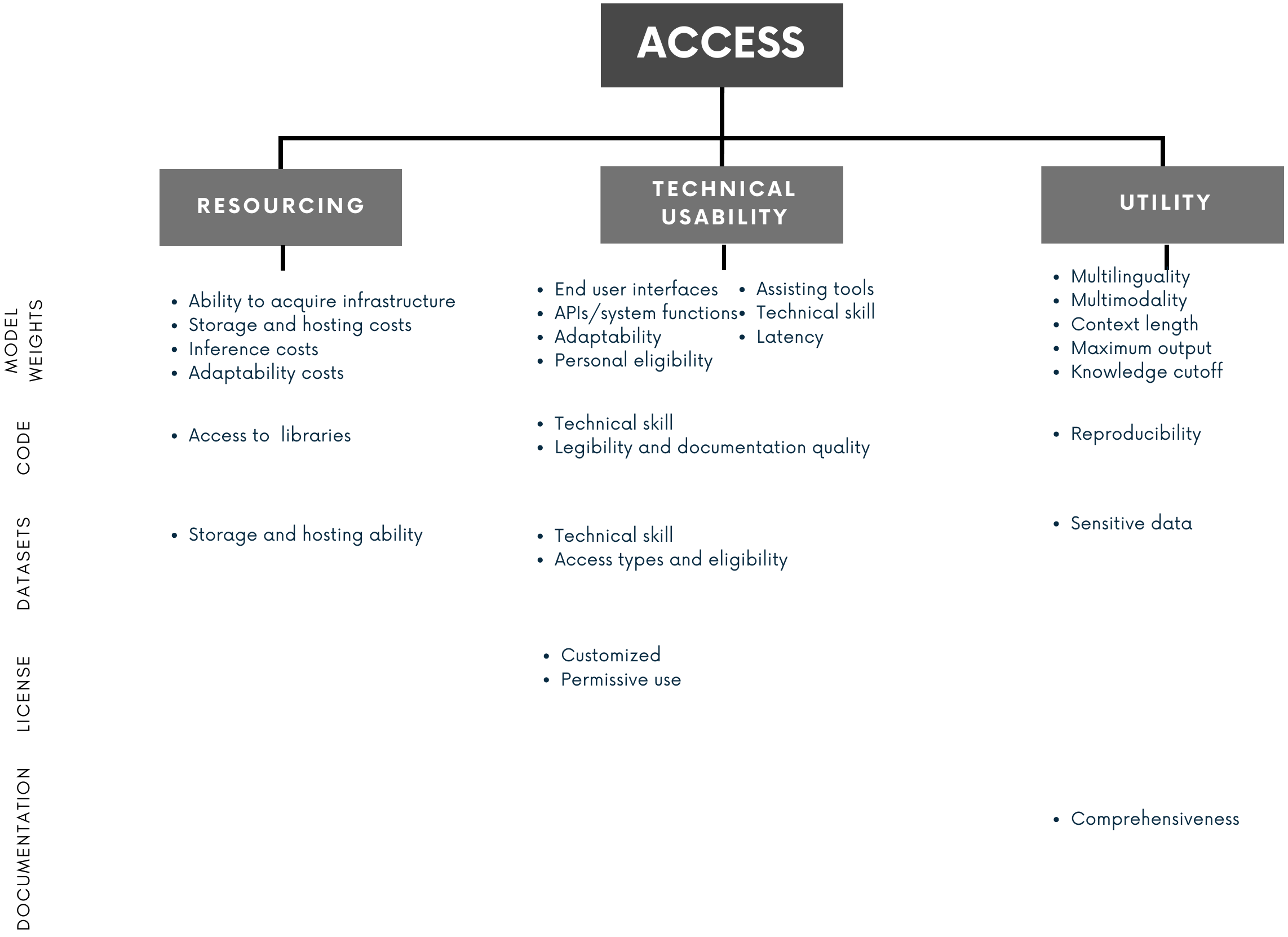}
  \caption{Categories of Access and Respective Variables}
  \label{fig:accesscat}
\end{figure}

We most closely examine further considerations for system accessibility. Who is granted access and how they are able to use a given component shapes who is able to benefit and risk for malicious use. In order to examine who can access which component, more granularity is needed. The three subsections of access are resourcing, usability, and utility, as shown in Figure \ref{fig:accesscat} with the respective variables per component. 

\pagebreak

While these three categories necessarily overlap, this framing better structures the core means of using a system. Resourcing applies most to hardware and monetary costs of accessing a system. Technical usability refers most to software features that enable usage. Utility covers qualities and capabilities such as language and modality of system components, which also overlaps with but is distinct from task performance. Some variables, such as ability to adapt or fine-tune, can apply to multiple categories (\nameref{Resourcing} and \nameref{Technical Usability}), but are sorted into the most applicable category. The three categories ensure that access is considered from three vantage points. 

Risks and benefits affect stakeholders throughout the deployment timeline, such as developers and deployers, third-party deployers who are not the developers, researchers, users and customers, policymakers and regulators, and members of society who are affected by AI outputs and distribution. Risks and benefits noted are non-exhasutive, and do repeat throughout variables and access categories. This repetition stresses how many access variables can affect accessibility and have similar outcomes. Listed risks and benefits are tied to the listed variable, not necessarily deployer decisions. Stakeholders affected are be different by variable and scenario; for example, some models’ inference cost may disproportionately affect researchers compared to third-party deployers. How tradeoffs are weighed will depend on deployers determining release and policymakers’ priorities.

In each section, we highlight system components, as defined in existing frameworks for an AI system \cite{basdevant_towards_2024}, namely product; model components of datasets, code, and model weights; infrastructure, documentation, and licensing. We compare the following highest performing language models: 

\begin{itemize}
  \item Llama 3.1 405B Instruct (Meta, open-weight)
  \item DeepSeek v3 (DeepSeek, open-weight)
  \item GPT-4 (OpenAI, closed-weight)\footnote{Note: GPT-4o surpasses GPT-4, but GPT-4 is selected for this paper as the most comparable model.}
  \item Claude 3.5 Sonnet (Anthropic, closed-weight)
\end{itemize}

We select these models based on publicly available data on evaluation results and comparative performance; they are the top scoring open- and closed-weight models on Stanford University’s HELM Lite leaderboard \cite{noauthor_holistic_2025} as of early 2025. Information about each model is selected from their respective documentation cards \cite{noauthor_meta-llamallama-31-405b-instruct-fp8_2024}, technical reports \cite{openai_gpt-4_2024, deepseek-ai_deepseek-v3_2025}, blog announcements \cite{anthropic_claude_2024}, and product pages \cite{noauthor_openai_2025, noauthor_models_2025}. 

\subsection{Resourcing}
\label{Resourcing}
Resourcing refers to if and how a broad and diverse population can host and serve a system and its components, particularly around infrastructure and costs. Computational infrastructure is a notable barrier; outside of training, AI systems require varying levels of compute for hosting, inference, and functions such as fine-tuning. Compute resources skew heavily toward industry organizations \cite{ahmed_growing_2023} as costs remain high. The high cost of resourcing has been scrutinized as reducing accessibility of open-weight systems \cite{widder_why_2024}.

Local hosting refers to hosting a model on independent, local hardware, self-hosting can refer to hosting a model locally or renting external cloud instances, and external or managed hosting is when another organization is responsible for infrastructure and hosting costs. External or managed hosting includes dedicated deployments with dedicated infrastructure and serverless deployments with shared infrastructure. Costs for locally and self-hosting can be ambiguous and highly dependent on how hardware and cloud service providers are chosen. Most popular models are available for free limited use via interface when externally hosted by the developer or a hosting platform \cite{noauthor_huggingchat_2025}.

\subsubsection{Model}
\hfill\\
\textbf{Ability to acquire infrastructure} includes hardware and local infrastructure as well as cloud compute credits needed to locally, self-host, or serve a model. The type of hardware needed will differ by model, with larger models running better on only graphics processing units (GPUs) than the more accessible central processing units (CPUs), and some smaller models running on GPU-CPU mixed infrastructure or even CPUs. The global GPU shortage and compute demand skews who is able to acquire hardware. Renting cloud instances is often a more accessible option. Developers and deployers often offer free credits (e.g. Google Cloud new customer credits, OpenAI Dev Day credits, HF ZeroGPU).

\textit{Benefits: }Acquiring infrastructure allows users to access benefits from local and self-hosting models. Free credits can help low-resource researchers and developers. 

\textit{Risks: }Infrastructural needs and related costs may change for newer models. Sectoral gaps in who is able to obtain compute resources can disadvantage groups such as academia. Free credits can be misused by malicious actors. 

\textbf{Storage and hosting costs }include costs to locally or self-host a model. For local hosting, this includes cost of hardware, storage capacity for storing and loading models, electricity costs, and operational costs such as setup and maintenance. Memory requirements depend on the model and hardware. High upfront setup costs for locally hosting often leads to preferencing self-hosting or using API alternatives. Self-hosting requires user accounts with cloud providers and hosting platforms.

\textit{Benefits: }Local or self hosting a model gives more controllability and privacy to the user, especially for sensitive data. Long-term usage costs can be lower, depending on the model and usage. Local hosting ensures stability of access and ensures access to the same version of the model, with no external dependencies. It also enables unrestricted fine-tuning and other adaptations for better customization. 

\textit{Risks: }Building and managing local infrastructure has a high upfront cost and necessitates hardware investment, in addition to the technical knowledge of setting up and optimizing resources. Energy costs may also be high, and add to potential costs for general operation, from human hours in setup and maintenance to cooling and security. Locally hosted models are less monitorable or not monitorable, especially compared to models hosted by deployer organizations; malicious actors can locally host a model and adapt or output harmful content while unmonitored.

\textbf{Inference costs} for locally or self-hosted models are determined based on cloud instance selected and instance cost per hour, whereas externally hosted models costs are often charged per token. Depending on infrastructure, locally or self-hosted models do not have limits and tokens can be generated as needed. For externally hosted models, this is usually measured in cost per token or output. Some high performance tasks aggregate multiple calls to a model \cite{chen_are_2024}, using multiple weaker model calls to match a stronger model, although research shows limitations to this approach \cite{stroebl_inference_2024}. Often model sizes correlate to popular hardware limitations for a node. Inference optimization methods can be conducted at the data, model, and hardware levels, and aim to make memory, storage, and computation more efficient. For popular decoder-only language models, continuously inputting lengthy token sequences is costly and is often mitigated using a key-value cache (KV cache) that stores and reuses previous key and value pairs \cite{zhou_survey_2024}. Compression techniques such as model quantization \cite{mao_compressibility_2024} can reduce the computational cost of inference, allowing models to run on less computing power. Other popular techniques include model distillation, where a smaller model is trained from a larger model to emulate behavior and performance, and sparsification, such as via pruning techniques, which remove some weights. Chosen techniques can result in performance losses \cite{zhu_survey_2024}, depend on tasks, and can be combined \cite{xu_survey_2023}.

\textit{Benefits: }Local or self-hosted task-specific small models can achieve high performance at relatively low cost while reducing latency \cite{irugalbandara_scaling_2024}. Per token or per output cost is easier for users without or unable to invest in local infrastructure, or short-term users. Costs for large generations or high quality outputs can disincentivize malicious actors. 

\textit{Risks: }Large and compute-intensive models are generally costly for inference and will have higher latency. High costs can disincentivize users and be inequitable for who is able to benefit. Powerful on-device models are hard to monitor.

\pagebreak

\textbf{Adaptability costs} include costs for fine-tuning or reinforcement learning with human feedback. In addition to computing power, human labor and time are often costly. As opposed to classic fine-tuning, narrow approaches to fine-tuning can be more efficient and less costly. Approaches such as parameter-efficient fine-tuning (PEFT) \cite{xu_parameter-efficient_2023}, from low-rank adaptation (LoRA) \cite{hu_lora_2021} to quantized low-rank adaptation (QLoRA) \cite{dettmers_qlora_2023}, can have performance trade-offs. See \textit{Ability to adapt or fine-tune} in \nameref{Technical Usability} for usage considerations. 

\textit{Benefits:} The relatively small amount of data and compute needed compared to training is more accessible. 

\textit{Risks: }Lowering the barrier to adapt models can enable harmful model behavior.

\subsubsection{Code}
\hfill\\
\textbf{Access to libraries} includes the ability for users to run code that includes libraries that may be only available to a certain organization.

\textit{Benefits: }Full access to all parts of the code can empower researchers to reproduce and build on existing work. Lack of access can prevent malicious actor reproduction.

\textit{Risks: }Without access to all libraries included, the code is largely unrunnable, preventing scientific reproduction. 

\subsubsection{Training Data}
\hfill\\
\textbf{Storage and hosting ability} refers to capacity to host datasets, especially large datasets and those with modalities with high storage requirements. 

\textit{Benefits:} Being able to store datasets such as training data enables additional research and the ability to improve data quality. The Pile dataset at 1346GB in practice is relatively accessible to host \cite{gao_pile_2020} and has contributed to many other datasets and models including LLaMA \cite{touvron_llama_2023}.

\textit{Risks:} For sensitive datasets, secure storage and sharing is needed to prevent misuse or malicious use. 

\subsubsection{Comparing models}
Comparing hosting and inference costs for open-weight models depends highly on utilization, task, latency, and hardware used, and varies heavily. Given numbers are broad approximations, and change with hardware, usage, and other factors. Precision affects cost, with many popular models trained with 32 or 16 bits of precision.  Memory needed for inference tends to be around 20\% higher than model memory \cite{anthony_transformer_2023}.  While hardware memory varies, we use NVIDIA H100 GPUs, which have 80 GB of VRAM, and NVIDIA H200 GPUs, which have 141 GB of HBM3e memory. These GPUs cost approximately \$25,000 and \$32,000 each, respectively. Costs for hardware, renting GPUs, energy, and production models fluctuate heavily over time. See Appendix \nameref{Resourcing Calculations}  for more information on determining resourcing for Llama 3.1 405B Instruct.

\textbf{Llama 3.1 405B Instruct} to run locally, uses at minimum 8 NVIDIA H100 GPUs and 405 GB VRAM to load the model in a lower 8-bit precision (FP8) and at least 810 GB VRAM and 11 H100 GPUs to load it in its original 16-bit precision (FP16). To reach the 128K token maximum context length, cache memory requirements are 123.05 GB, which would bring hardware requirements to 12 H100 GPUs. Full fine-tuning requires 3.25 TB memory \cite{noauthor_meta-llamallama-31-405b-instruct-fp8_2024}. This results in an upfront cost of at least \$200,000 for GPUs in FP8 and \$300,000 in FP16, plus additional costs for servers. 

\textbf{DeepSeek v3} to run locally is recommended to use 8 NVIDIA H200 GPUs to deploy in FP8. The model is trained in FP8, but can be run in BF16 \cite{deepseek-ai_deepseek-v3_2025}. Due to its mixed precision training framework, the model is relatively small for its performance, at about 1.3 TB. To deploy the BF16 variant doubles memory requirements to 2.5 TB \cite{poulopoulos_deploying_2025}. At minimum, base infrastructure costs \$256,000, plus costs for servers. Distilled versions of DeepSeek’s reasoning model, DeepSeek-R1, which is trained on DeepSeek v3, boast relatively high performance and can run on one GPU with small models, such as the 8B parameter model, able to run on a personal computer 
\cite{pagezy_how_2025}.


\textbf{GPT-4} is available via the OpenAI API, where the developer carries storage, hosting, and maintenance costs. As a closed-weight model, it is not possible to host locally or self-host. Information on hosting costs is unavailable publicly. 

\textbf{Claude 3.5 Sonnet} is available via the Anthropic API, with the developer carrying related hosting costs. It is also closed-weight without publicly available hosting information.

\resizebox{\textwidth}{!}{
\rowcolors{2}{gray!25}{white}
\begin{tabular}{ |p{1.5cm}||p{4.2cm}|p{4.2cm}|p{4.2cm}| p{4.2cm}| }
 \hline
 \multicolumn{5}{|c|}{Resourcing: Comparing Models} \\
 \hline
  &Llama 3.1 &DeepSeek v3 &GPT-4 &Claude 3.5 Sonnet\\
 \hline
 Hosting Infra-structure   & 12 H100 GPUs, Server, \par Maintenance &8 H200 GPUs, Server, \par Maintenance &   N/A (deployer supplied) &N/A (deployer supplied)\\
 Hosting Cost &   12 H100 GPUs, Energy, \par Maintenance  & 8 H200 GPUs, Server, \par Energy, Maintenance   &N/A (deployer supplied) &N/A (deployer supplied)\\
 Inference Cost &(12 H100 GPUs \par x GPU / hour) /response rate & (8 H200 GPUs \par x GPU / hour) /response rate &  N/A (deployer supplied) &N/A (deployer supplied)\\
 API Cost Per Token    &Input: \$5 / 1M T
\par Output: \$16 / 1M T &  \$0.07 (cache hit) or \$0.27 (cache miss) / 1M T \par Output: \$1.10 / 1M T & Input: \$10 / 1M T \par Output: \$30 / 1M T &Input: \$3 / 1M T 
\par Output: \$15 / 1M T\\
 \hline
\end{tabular}
}

\subsection{Technical Usability}
\label{Technical Usability}
Usability determines whether and how a broad and diverse population can technically access and use an AI system and its components. For an average user seeking to query a model, well-designed and easy user interfaces have enormously lowered the barrier for people outside the AI field to engage with these systems. However, reaching more people also carries a risk of enabling malicious uses from a larger set of users. 

\subsubsection{Model}
\hfill\\
\textbf{End user interfaces}, including interfaces to connect inputs and outputs and chat with a model, lower the barrier to interact with a model and allow broader populations to access a model. These can be built and maintained by hosting organizations, and projects \cite{oobabooga_oobaboogatext-generation-webui_2024} with open repositories can ease interface building for locally hosted models. Interfaces are also built for tasks such as training, fine-tuning \cite{asaria_transformer_2023}, and evaluating models, also lowering those barriers.

\textit{Benefit: }Well-designed interfaces surpass the need for technical expertise, and enable users without a technical background to complete technical tasks \cite{school_redefining_2023}. These interfaces can make models more enjoyable to use by reducing friction, and help users more easily understand how to effectively interact with a model. Effective interfaces, user experience, and design can help models such as ChatGPT appeal to wider audiences \cite{tolhurst_how_2023}. 

\textit{Risk:} While interfaces enable subject matter experts to use AI, they can also enable less technical malicious actors to generate harmful material.  Popular interfaces have been used to consult in terrorist attacks \cite{cameron_before_2025, singleton_how_2023}. Early interfaces for Bard and ChatGPT were used to generate malicious code \cite{pearl_googles_2023}. Deepfake applications that build an application-specific interface on top of image generative models have been weaponized for non-consensual intimate imagery (NCII) and used by and against minors \cite{north_ai_2024}. Additionally, without full transparency, an interface can obfuscate potential layers on top of a model that a user could be interacting with. This includes content filters which can skew perception of model performance. Interfaces also can often not be built upon. Underlying models can change without notice. 

\textbf{AI system functions such as remote APIs} can include logit access. APIs bridge systems, providing easier model access to builders. Model releases, whether open- or closed-weight, may or may not also have an accompanying API hosted by the developer. The level of access that an API provides, such as fine-tuning, can vary as well.

\textit{Benefits: }APIs enable users to call a model and build more easily, streamlining and simplifying model integration, and helping projects scale easier. Remote APIs also can be monitored and rate limited, providing a means of intervention in deployment. Users can be authenticated and blocked at any time. Misuse can be shut down \cite{orland_viral_2025}.

\textit{Risks:} APIs can introduce security vulnerabilities such as stolen or leaked API keys, and can be a vector for attacks such as stealing model information \cite{carlini_stealing_2024, finlayson_logits_2024} or distributed denial of service attacks. Sensitive data and sensitive inputs can also be leaked. API stability can vary by provider and high demand can slow response rates. Models and versions available by API can change without notice. Concerns around API monitoring include surveillance and privacy, especially for inputting sensitive or proprietary data. 

\textbf{Ability to adapt or fine-tune} grants access to adapt a model toward a set of determined tasks or behaviors. This can be enabled via API. Adapting models is possible for open-weight models for those with adequate compute and technical skills, and can be provided for hosted, closed-weight models.

\textit{Benefits:} Adapting or fine-tuning a model allows users to tailor models to use cases, products, and contexts. Customizability can improve models for specific use cases.

\textit{Risks:} Fine-tuning can override safety techniques and guardrails even in closed-weight but fine-tunable models \cite{henderson_safety_2024}, even without malicious intent \cite{qi_fine-tuning_2023}. 

\textbf{Personal eligibility }includes the user age, countries, regions, and other restrictions that may be part of a model and system’s availability. This can be limited by gating or simply blocking IPs from unsupported regions. Restrictions can be technically implemented or legally such as with terms of use.

\textit{Benefits: }Age restrictions can prevent minors from inappropriate content. Regional access restrictions can prevent misuse in or from those regions. For platforms that deploy or host models, user accounts can aid in monitoring.

\textit{Risks: }Collecting user data for permissions can raise both verification issues and privacy concerns. Blocking usage in some regions can prevent populations from benefiting and can lead to technological skill asymmetry. Verification for legal mechanisms vary in reliability and terms of use can be difficult to enforce. 

\textbf{Assisting tools, libraries, courses, and frameworks} aid in running, deploying, and adapting models. This includes features and prewritten code on collaborative platforms like Hugging Face. Examples of libraries are PyTorch and Tensorflow, frameworks include Ollama, LangChain, and LlamaIndex.

\textit{Benefits:} Many free and available resources train users and simplify deploying pre-trained models. This can close sectoral and resource gaps, especially for underrepresented groups, low-resource developers, and startups. 

\textit{Risks:} Malicious actors have equal access to open tools.

\textbf{Technical skill }refers to the expertise needed to run, host, adapt, conduct research on, and deploy a model. 

\textit{Benefits:} Layperson attackers will be unable to conduct sophisticated attacks without upskilling. Running open weight models will require understanding the needed infrastructure.  Conducting attacks and and surpassing model safeguards \cite{rando_red-teaming_2022} requires knowing potential existing work \cite{chao_jailbreaking_2024} and proven techniques. 

\textit{Risks: }The technical proficiency needed to meaningfully work with a model limits who is able to benefit. Educational resources can also aid attackers seeking to upskill. Open-weight models without existing infrastructure such as an interface or API require technical knowledge to deploy, from setting up the right environment to configuring a model. 

\textbf{Latency} refers to the time needed to process inputs and generate an output. Factors such as infrastructure, model capability, and input size and complexity can affect latency. 

\textit{Benefits:} Locally hosting models can reduce latency and model usage friction, improving user experience \cite{lee_evaluating_2024}. 

\textit{Risks: }Powerful models hosted on external infrastructure, such as via API or hosted interfaces, can face high latency with high demands to their infrastructure. This can make building beneficial projects harder. Low-latency, especially for locally hosted models, can allow malicious actors to generate larger volumes of harmful content faster.

\subsubsection{Code}
\hfill\\
\textbf{Technical skill }includes the ability to run code accompanying system releases. It is related to \textbf{legibility and quality of documentation}, which refers to how easily code that accompanies models can be understood and used. Legacy code can become too convoluted for easy use by external actors. 

\textit{Benefits:} High quality code release better incentivizes reuse. 

\textit{Risks:} Poor quality code can make additional research more difficult, with barriers to onboarding researchers and parsing past decisions.

\subsubsection{Training Data}
\hfill\\
\textbf{Technical skill} refers to the ability to conduct research and analysis on datasets, in addition to using datasets accompanying system releases for development.

\textit{Benefits:} Data research can give insights into model performance and safety. 

\textit{Risks: }Datasets can contain harmful and illegal material.

\textbf{Access types and eligibility} to access datasets is similar to that for models, such as restrictions by user age and region. It can be further affected by legal constraints, from intellectual property (IP) and personally identifiable information (PII) access. Types of access include direct access, API access to certain functions such as the ROOTS tool \cite{piktus_roots_2023}, API access to statistics about data, legal (e.g. GDPR) requests, and court discovery. 

\textit{Benefits:} Limiting dataset access can prevent unauthorized users, such as minors, from accessing inappropriate material. Legal compliance for sensitive data protects data subjects. 

\textit{Risks: }Restricting access to datasets limits who can conduct research.

\subsubsection{License}
\hfill\\
\textbf{Customized} licenses can range from permissive use for researchers but restricted for commercial applications. \textbf{Permissive use} as allowed by licenses for models, code, and data \cite{white_model_2024}, with the types of licenses and applicability differing by system component. For models, use may not be tightly correlated with weights specifically and instead with the system usage overall. For data, rights attached to data are an important factor. 

\textit{Benefits: }Allowing access to researchers can enhance scientific integrity. Customized licenses can vary with size of a model; permissive licenses can be staged to allow open usage for smaller parameter count models and tailored, more restrictive licenses for larger models, as seen with Alibaba's Qwen2 model family \cite{yang_qwen2_2024}. Permissive licenses can grant users security. 

\textit{Risks: }Licenses, especially when customized, can be difficult to understand for an average user with limited legal experience. Increasingly customized licenses are often more complex legally and can be difficult to enforce. 

\subsubsection{Comparing models}

Many popular models will have some level of free interface to generate outputs, often with limitations (rate, filters), and hosted by the developer or third party sources. Latency will vary, depending on infrastructure and setup for open-weight self-hosted models, and depending on demand for API-served models. 

\resizebox{\textwidth}{!}{
\rowcolors{2}{gray!25}{white}
\begin{tabular}{ |p{2.5cm}||p{2.7cm}|p{2.7cm}|p{2.7cm}| p{2.7cm}| }
 \hline
 \multicolumn{5}{|c|}{Technical Usability: Comparing Models} \\
 \hline
  &Llama 3.1 &DeepSeek v3 &GPT-4 &Claude 3.5 Sonnet\\
 \hline
 User Interface   & Yes: Developer, Third-Party    &Yes: Developer, Third-Party & Yes: Developer &Yes: Developer\\
 API &   Yes: Third-Party  & Yes: Developer, Third-Party   &Yes: Developer, Third-Party &Yes: Developer, Third-Party\\
 Ability to Fine-Tune &Yes & Yes &  Yes &Yes\\
 License    &Customized & Customized & Customized &Customized\\
 \hline
\end{tabular}
}
\subsection{Utility} 
\label{Utility}
Where usability refers to whether a broad and diverse population can technically access a system, utility refers to whether populations can gain utility from accessible capabilities of the system. This does not cover overall or task-specific model performance, which is noted in \nameref{Non-Access Considerations}.

\subsubsection{Model} 
\hfill\\
\textbf{Multilingualilty} considers high quality outputs in multiple languages.

\textit{Benefits: }High quality language performance, especially for non-English and low-resource language speakers, opens new markets to AI and allows native speakers of many languages \cite{adelani_masakhaner_2021, tsanni_this_2023} to more comfortably use, build, and deploy.

\textit{Risks: }More language availability and quality gives access to attackers in that language and in regions where that language is more popularly spoken. Some languages may be more spoken in areas with lower resourcing for monitoring or interventions. This disparate resourcing has proved dire for social media monitoring failures \cite{stecklow_why_2018}.

\textbf{Multimodality }covers the modalities for model inputs and outputs, such as text and code, image, audio, and video. 

\textit{Benefits:} Certain modalities, e.g. text, fit better for given use cases and deployment contexts, e.g. summarization.

\textit{Risks:} Different modalities require different approaches to monitoring and safeguarding. Some modalities are more prone to certain attacks than others, such as images for NCII generation.

\textbf{Context length} is the possible size of an input to a model.

\textit{Benefits:} Larger input capacities can improve output relevance, especially for inputs with large spread or complex information. Long context lengths can enable more use cases and improve output accuracy and user experience. 

\textit{Risk: }Long context windows can be exploited to jailbreak LLMs \cite{anil_many-shot_nodate} and can be more expensive to use. 

\textbf{Maximum output }refers to the possible size of the output generated, often measured in tokens. 

\textit{Benefits:} Longer outputs can improve some use cases. 

\textit{Risks: }Longer outputs can aid in large-scale attacks that require more content, such as spreading disinformation.

\textbf{Knowledge cutoff} is the timeframe from when training data is stopped. 

\textit{Benefits:} Later knowledge cutoffs can provide better updated results in modern contexts. 

\textit{Risks:} Earlier model cutoffs can lead to inaccurate or misleading results for current events. 

\subsubsection{Code}
\hfill\\
\textbf{Reproducibility} includes sharing code for training models and other tasks. Comparisons between open systems and open software continue to influence tradeoff analysis \cite{langenkamp_how_2022}.

\textit{Benefit:} Code release has proven helpful scientifically for reproducibility and collaboration \cite{pineau_improving_2021}, leading to higher citations for publications with accompanying code \cite{zhou_what_2024}. 

\textit{Risk: }Malicious actors can use code to reproduce components.

\pagebreak

\subsubsection{Training data}
\hfill\\
\textbf{Sensitive data} includes IP and copyrighted material as well as private and PII.

\textit{Benefits:} For some use cases in controlled settings with appropriate compliance, training on sensitive data can produce helpful and tailored outcomes. 

\textit{Risks: }Including sensitive data in training data risks leakage, stealing, and harm to data subjects. 

\subsubsection{Documentation}
\hfill\\
\textbf{Comprehensiveness} is the robustness and legibility of documentation for system components and processes. Documentation can give insights into system components such as from where public training data was sourced. In addition to documentation differing in findability and consumability, many processes and aspects of model training and release are not documented or shared publicly; for example release decisions processes are rarely if ever shared \cite{bommasani_foundation_2023}.

\textit{Benefits: }More comprehensive documentation that includes information such as evaluation results and uncertainty increases transparency \cite{liao_ai_2024} and trust. Useful details in technical papers increase reproducibility for scientific integrity.

\textit{Risks: }Incomplete dataset documentation or documentation with sparse sections hurts documentation quality \cite{yang_navigating_2024}. The same details in technical papers can be misused by malicious actors to reproduce components.

\subsubsection{Comparing models}
Utility across the four selected models are largely comparable, with strong multilinguality availability in  DeepSeek v3. Notable differences between the below open- and closed-weight models are in input modality, where GPT-4 and Claude 3.5 Sonnet are vision language models and allow image inputs.

\resizebox{17cm}{!}{
\rowcolors{2}{gray!25}{white}
\begin{tabular}{ |p{2.5cm}||p{5.5cm}|p{5.5cm}|p{5.5cm}| p{5.5cm}| }
 \hline
 \multicolumn{5}{|c|}{Utility: Comparing Models} \\
 \hline
  &Llama 3.1 &DeepSeek v3 &GPT-4 &Claude 3.5 Sonnet\\
 \hline
 Multilingual   & English, German, French, Italian, Portuguese, Hindi, Spanish, Thai    &English, Afrikaans, Albanian, Amharic, Arabic, Armenian, Azerbaijani, Basque, Belarusian, Bengali, Bulgarian, Burmese, Cambodian, Catalan, Croatian, Czech, Danish, Dutch, Estonian, Filipino, Finnish, French, Galician, Georgian, German, Greek, Gujarati, Hebrew, Hindi, Hungarian, Icelandic, Indonesian, Italian, Japanese, Kannada, Kazakh, Korean, Kyrgyz, Laotian, Latvian, Lithuanian, Macedonian, Malay, Malayalam, Marathi, Mongolian, Nepali, Norwegian Bokmål, Persian, Polish, Portuguese, Punjabi, Rhaeto-Romance, Romanian, Russian, Serbian, Simplified Chinese, Singhalese, Slovak, Slovenian, Spanish, Swahili, Swedish, Tamil, Telugu, Thai, Traditional Chinese, Turkish, Ukrainian, Urdu, Vietnamese, Zulu & English, Italian, Afrikaans, Spanish, German, French, Indonesian, Russian, Polish, Ukrainian, Greek, Latvian, Mandarin, Arabic, Turkish, Japanese, Swahili, Welsh, Korean, Icelandic, Bengali, Urdu, Nepali, Thai, Punjabi, Marathi, Telugu &English, German, Spanish, French, Hindi, Italian, Japanese, Korean, Polish, Portuguese, Russian, Turkish, Chinese \\
 Modalities & Input: Text, Code Output: Text, Code  & Input: Text, Code Output: Text, Code &Input: Image, Text, Code 
 Output: Text, Code &Input: Image, Text, Code 
 Output: Text, Code\\
 Context Length (In Tokens) &128k & 128k &  128k &200k\\
 Maximum Output Tokens    &4096 & 8000 & 4096 &8192\\
 Knowledge Cutoff    &December 2023  & October 2023 & December 2023  &April 2024\\
 \hline
\end{tabular}
}

\pagebreak

\section{Scaling Access: Access to Whom}
\begin{figure}[h!]
  \centering
  \includegraphics[width=\textwidth]{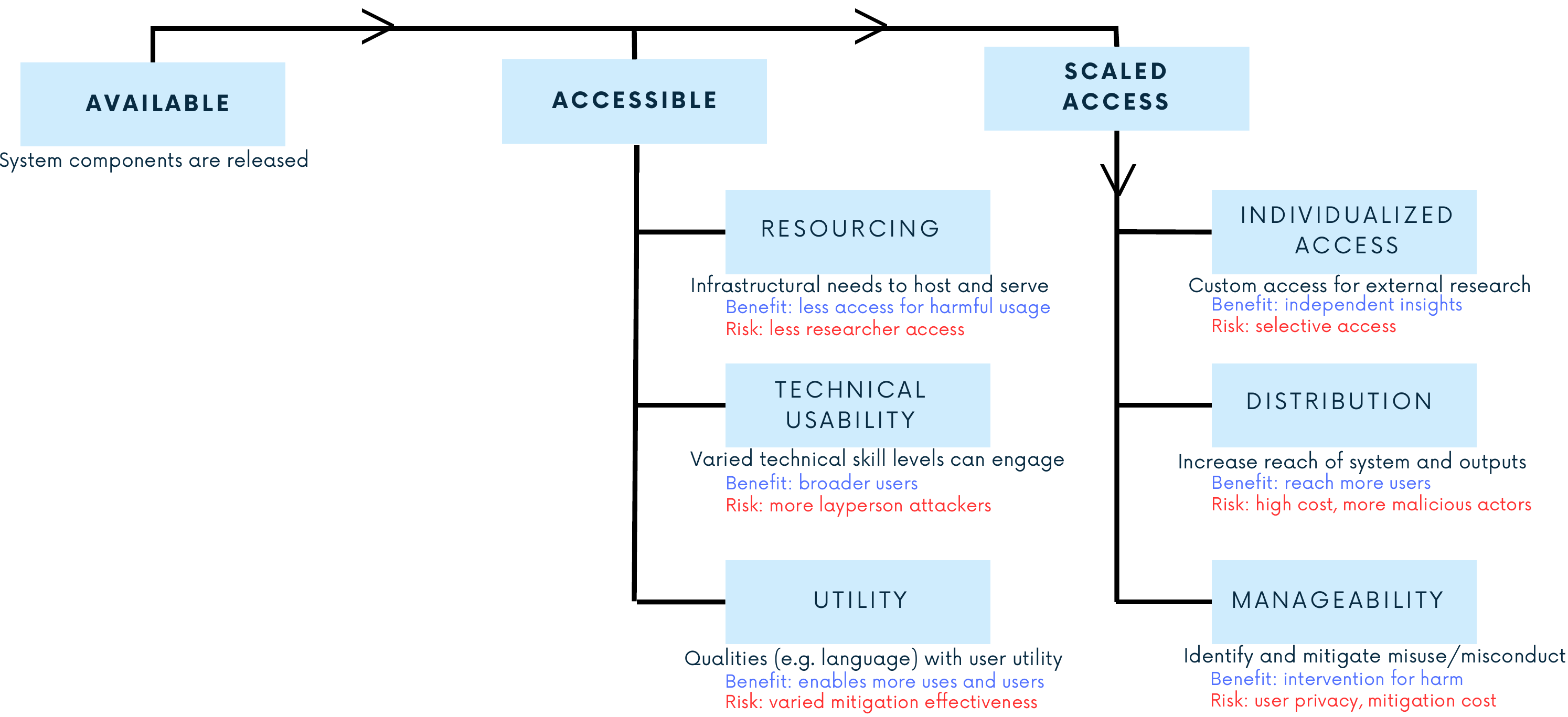}
  \caption{Flow of Access and Scale with High Level Tradeoffs}
  \label{fig:scaleflow}
\end{figure}

Who can and should be able to access components often centers on motivation. Some model licenses allow research access and not commercial deployment, while some models are broadly open to any usage. Additionally, the ability to store, host, and deploy artifacts may differ by factors such as geographic location. For sensitive artifacts such as some datasets, the ability to accommodate regional laws \cite{jernite_data_2022} is crucial. As systems gain reach, the scale of who can access a model has a positive relationship with both who can benefit, and potential malicious actors. The scale also influences deployer organizations’ ability to manage risk and intervene in violations. Making system components accessible sets the foundation for scaling who is given access. Increased access will add computing power demands and must consider means of distribution. This can affect variables such as latency. Depending on the diversity of users, usability and utility variables may update as well, such as increasing language compatibility when deploying in new regions. Figure \ref{fig:scaleflow} shows the flow from release, to becoming accessible, to scale considerations.

\subsection{Individualized Release and Depth of Access}
One of the central questions of access is which actors qualify for access, especially in limited releases. Calls for independent scrutiny of closed systems, such as structured access \cite{shevlane_structured_2024}, safe harbor evaluations \cite{longpre_safe_2024}, and third party audits \cite{noauthor_how_2023}, must ask what components an external actor has access to, and what is the permissive ability to access each component. Mostly targeted toward researchers and auditors, depth of access enables scrutiny outside of the developer’s expertise or incentives. For product and commercial-oriented actors, depth enables better tailored models for a given application. Per system component, the amount of access, or the depth, given to an external actor, researcher, or auditor depends on the use case and research area \cite{bucknall_structured_2023}. The depth of access overlaps with what is made available, and is sometimes framed as “completeness”. Some research often requires model weights, such as interpretability \cite{bereska_mechanistic_2024}.

Safe information sharing mechanisms can encourage collaboration and transparency \cite{trask_beyond_2024}. Some data may not be fit for public release, often due to legal or security constraints, such as trade secrets, IP, PII, and national security information. Existing frameworks, such as the Five Safes, help manage safe research access to data \cite{service_what_nodate}. 

For systems developed with researcher access, more work is needed to understand the most urgent types of research that should be conducted on that system and the correlating level of access needed. Curated external access to a closed-weight system faces challenges of appropriate due diligence for determining the right users to whom access should be granted. Curating access by select priority expertise areas may overlook potential unforeseen contributions. 

\subsection{Release Distribution Methods}
Who can and will access a system and its components is tied to how a system is distributed. Distribution considers the method of distribution and the means by which components are made available. This can include developer and deployer platforms, hosting platforms, curated access to individuals, social media links, and more rudimentary methods such as USB sticks or QR codes \cite{r0b0tsp1der_yacinemtb_2023}. Staging releases with a set time frame for making model weight available \cite{solaiman_release_2019} can aid both understanding of threats and assuring trust in the release process. Developer and deploy platforms can advertise, influencing who is incentivized to access the platform. Platforms that host model weights can include the model developer’s sites, hosting platforms like Hugging Face, and partner platforms. For example, Llama models are available on Meta official websites, Hugging Face repositories, Kaggle, and partners for different model sizes \cite{noauthor_llama_2024}. Hosting platforms can enable broad access to systems, which can be controlled and managed via gating and user access. Varied platform visibility for proprietary, private, or sensitive systems can allow organizations to selectively share components, such as internal only organizational access. Content guidelines and terms of use are applied at the model level. 

\textbf{Distribution of content }affects how impactful a model and its outputs are, for beneficial and malicious purposes. For example, AI generated disinformation is only as impactful as its distribution, with studies showing concerns about AI disinformation are overstated \cite{simon_misinformation_2023}. Concerns can of course change with time and major events, such as elections, with distribution remaining a key factor \cite{solaiman_five_2024}. The perception of heavy distribution can have harmful effects: a survey of 100,000 people across 47 countries showed 59\% of people are worried about false news content \cite{newman_reuters_2024}. Integrating AI into distribution platforms, such as news media, has been met with public skepticism: the majority of respondents in the same survey in the U.S. and UK noted they would be uncomfortable with news produced mostly with AI \cite{newman_reuters_2024}.

\textbf{Scale of deploying systems }requires heavy investment in maintaining the means of deployment, often an API, user interface, or other application. From rapid adoption by daily users \cite{mcclain_americans_2024}, to increasing usage by researchers \cite{liang_mapping_2024}, to deployment by startups and businesses \cite{noauthor_state_2024}, reaching more people via deployment also means meeting infrastructure, legal, and management demands. These demands tend to differ by user group. 

\subsection{Managing Scale}
Manageability includes the ability to establish what constitutes misuse and identify misconduct; to monitor and intervene on misconduct; to reduce the reach of a system; and the cost to manage. This can mean managing the release of artifacts themselves as well as managing deployed systems, especially when models are deployed by third-party commercial entities. The distribution of a release and breadth of access affects how monitorable and manageable a system is: a widely, globally used system across release availability will have different demands as user scale increases. Use in different e.g. languages can affect manageability. Misuse risks may not be fully addressed in either open-weight or closed-weight settings. Access controls largely depend on method of release, such as the ability to create an account to download model weights for open-weight releases or adhering to API policies for API-deployed closed-weight models. 

Open-weight releases can be downloaded and kept by users with the appropriate hardware, leading to inability to monitor if run locally. Reducing reach by deprecating model weights from distribution and hosting platforms \cite{noauthor_ykilchergpt-4chan_2022} can limit malicious actor access. Proposed methods for preventing harmful use via access span from structural, such as access control, to technical, such as task blocking \cite{henderson_self-destructing_2023}. When models are deployed via API or a hosted medium, technical interventions can be ineffective or the deployer organization must be able to meet the demands of high user traffic. This can have high management costs. Interventions such as query refusals and false positives in content filters can disincentivize users from engaging with that model or deployer. Variables that affect management include competitive pressure towards faster deployment, wider access, and less friction for users. 

\pagebreak

\section{Access-Adjacent Considerations}
In addition to the risk-benefit tradeoffs for variables under each access category, overarching factors post-release strongly influence threat and utility landscapes. How and with what intentions a system is deployed influences reach and subsequently system risks-benefit tradeoffs. Accessibility is linked to deployability; whether a commercial entity is able to technically and legally deploy a system depends on variables such as licensing and other governance mechanisms. 

\subsection{Changes Over Time}
\label{Changes Over Time}
The rapid pace of change in the AI landscape shifts the weight of some variables. Notably, changes in capabilities \cite{anderljung_frontier_2023}, cost \cite{cottier_rising_2024}, and data availability \cite{longpre_consent_2024} are some of the most influential recent factors. Emergent abilities \cite{wei_emergent_2022} and how to measure and validate them \cite{schaeffer_are_2023} raise research questions on AI system future capabilities and risks. 

Advancement and competitive pressure from releases such as DeepSeek-R1, showcasing low-cost, open-weight, high performance, are already shifting potential for all three access categories. Within weeks, releases such as Mistral Small 3 \cite{noauthor_mistral_2025} contribute to trends towards smaller models. The Allen Institute for AI (AI2) model T{\"u}lu 3 405B model \cite{lambert_tulu_2024} quickly boasted high benchmarks but also shared lessons on emphasis on reinforcement learning. 

Beyond overall system performance increasing over time, some aspects of model components have improved. Tokenization leaps for non-English languages has increased language availability, although pricing disparities remain \cite{ahia_all_2023}. Cohere’s Command R+’s tokenizer compressed non-English language text better than comparable models at its time of release \cite{ruder_command_2024}, and models such as DeepSeek v3 \cite{deepseek-ai_deepseek-v3_2025} show significant improvements in non-English and low resource languages, especially compared to previous models. Changes in hardware and hardware costs include advances in memory capacity, better computational price-performance ratio, and also potential walls \cite{hobbhahn_trends_2023}. 

Inference cost for existing and newer models can change \cite{chen_frugalgpt_2023} with commercial needs and energy limitations. Popularization of reasoning models can put more computing pressure on inference, where thinking tokens are more demanded. Decline in data availability affects the utility of data collection over time, which can advantage organizations with better access to data resources.

\subsection{Modality Influence}
As referenced in \nameref{Utility}, modalities for model inputs, model outputs, and datasets impact threat models. Some modalities have higher malicious use potential and comparably less research on safeguards. For example, voice cloning can be helpful in medical settings for patients who have lost their voices to disease \cite{morris_patients_2023}, but raises serious misuse concerns such as scams and fraud, prompting government response \cite{noauthor_approaches_2024}. Modalities that can more closely infringe on personal likeness, such as audio and image, navigate safeguards for consent. Many audio generation platforms for voice cloning restrict users from usage without consent of the data subject, but consent mechanisms are flawed and most do not require proof of consent \cite{rose_ai_2024}. Safeguard effectiveness differs by modality as well, as seen with watermarking and content moderation. While some modalities have more flexible use cases and some product use cases only function with certain modalities, some modalities are more conducive to certain attacks. All modalities can be weaponized for egregious misuse such as child sexual abuse material (CSAM), and some modalities such as image are used more often \cite{thorn_safety_2024}. Some modalities are better suited for distribution per distribution platform, such as Instagram primarily distributing images.

\subsection{Smaller Models}
In addition to independent projects that enable users to run models locally, such as llamafile \cite{hood_introducing_2023}, larger model families often offer small model options. High costs for larger models  incentivize research to maintain high performance in smaller models, as exemplified by models such as DeepSeek-R1, Qwen2 72B Instruct, Llama 3 70B, and Mistral Large 2 128B. 

\textbf{On-device deployment }means models can be hosted without cloud servers and on everyday hardware such as laptops and phones. Apple’s OpenELM \cite{mehta_openelm_2024} models ranging from 270M to 3B parameters, can run on Macbooks and some iPhones. Microsoft’s Phi models \cite{abdin_phi-3_2024} released a 3.8B parameter model, phi-3-mini, that can run on a phone and boasts comparable benchmark results to GPT-3.5. On-device hosting sidesteps the security risks associated with using the internet. Additional models in this category are Google’s Gemma and Mistral’s Les Ministraux model families.

\subsection{Application and Actualization}
A model’s usefulness or harmfulness is directly informed by the application in which it is deployed. Due to many generative systems’ generality in task performance, general risk assessments are unable to capture all possible risks and magnitudes of risks. Assessing model safety is best done in the context of the operational domain \cite{khlaaf_toward_2023}. Building application-specific infrastructure and optimizing for usability can be weaponized, as seen with deepfake NCII applications \cite{north_ai_2024}. Application-specific infrastructure can also be misused, where AI tools are integrated into platforms such as animation and modeling interfaces \cite{wiggers_uk_2024}. For certain risks, such as chemical and biological attacks, material and physical needs and equipment are needed to actualize harm. Creating harmful toxins and pathogens relies on the feasibility, both operationally and biologically, of developing a high risk attack \cite{mouton_operational_2024}. For high risk threats, effective threat modeling should consider how operationalizable, targeted, and scalable, attacks can be.

\subsection{Non-Access Considerations}
\label{Non-Access Considerations}
Some release considerations are adjacent to but not directly related to model accessibility. This includes overall and task-specific performance, scientific and market impacts, and usefulness outside of access variables such as agentic behavior \cite{shavit_practices_2023}. As discussed in \nameref{Changes Over Time}, capability increases are often centered in risk discourse, with advances with scaling \cite{kaplan_scaling_2020} informing some deployment decisions \cite{noauthor_responsible_2023}. High performance for high risk tasks, whether from a large generalized model or task-specific model, can include output quality. Output quality can range based on task. The realism and human-like quality of outputs affects usefulness. AI outputs close to or indistinguishable from human-generated outputs that are not labeled or shared as AI outputs contribute to clouding our information ecosystem. Task-specific performance, especially for high risk tasks, and training on dangerous information such as weapons manufacturing, add to potential threats. Mitigating these threats includes sorting data and redacting dangerous information or blacklisting potentially dangerous data sources. 

Scientific and economic factors are popularly referenced in release decisions, with themes of broader community feedback and the need for openness in scientific integrity and innovation \cite{society_science_2024}. Releasing training techiniques, as with DeepSeek-R1 \cite{deepseek-ai_deepseek-r1_2025}, can influence broader investment in for example, reinforcement learning.

Concentration of power \cite{thun_stopping_2024} in the market and research field is a commonly cited concern for who is able to access and build systems. This includes sectoral compute gaps and the ability for lower resource actors to compete commercially. Unexpected risks and ongoing safety research areas such as deception capabilities \cite{park_ai_2024} also affect release decisions. A related variable is the state of tamper-proof or tamper resistant safeguards \cite{tamirisa_tamper-resistant_2024} per modality, such as image watermarking \cite{an_waves_2024}. In addition to robustness, public trust in safeguards affects release receptivity.

\section{Limitations and Future Work}
While access considerations broaden insights into release, tangential artifacts and variables as well as non-access variables require more scrutiny. 

\textbf{System-adjacent components} that are not inherently part of development may not be included in model releases. Evaluations and test datasets, which can be developed by model developers or other research groups, can be withheld out of concern for contaminating training data \cite{elangovan_memorization_2021}. Model releases that do include evaluations include AI2's OLMo \cite{noauthor_olmo_2024}, where the code and data used to produce OLMo 2’s results are on the model release page.  

\textbf{Modalities with comparably less research background} and less access demand may have risks and benefits that evolve differently than other modalities. Incentives to reduce prices for popular commercial modalities such as text may differ from video. The limited data and publicly available documentation on some modality releases, such as audio and video, make  effective release strategies difficult to meaningfully compare.

\textbf{The pace of advancements and ecosystem influence }shifts costs and incentives in timeframes that are difficult to reliably model. Pricing changes as new model service providers enter the market leads to competitive pricing. Related to \nameref{Changes Over Time}, and similar to demands by modalities differing, ecosystem variables such as demand for access and economic influences in willingness to adjust to costs \cite{lee_how_2024}. Low-cost, open-weight models are more easily able to long-term embed in digital infrastructure, which can encourage broader integration due to version stability. As influenced by DeepSeek-R1's trend of high performance in low-cost, small models, examining potential thresholds for preferencing open-weight model integration is a critical open question that affects global and geopolitical influence. 

\textbf{Differentiating research versus commercial access} can be blurry; research access can inform and evolve into product and commercial ideas. More insight is needed on potentially distinctive uses and needs by sector.

\section{Conclusion}
Aspects of access beyond system release provide more clarity to release benefits, risks, and tradeoffs. Examining system accessibility and possible scale of access can more accurately inform release and policy decisions and research on release. We show through granular analysis and model comparisons that similar considerations exist across the release spectrum and for both open- and closed-weight models. We also show barriers without usable resources and infrastructure, and high risk via accessible interfaces and fine-tuning. As accessibility positively impacts scale, reaching more users broadens usage but also reaches more malicious actors and affects ability to intervene. The resourcing, technical usability, and utility subsections of access better exemplify how to more meaningfully weigh system usefulness and threats. 

\begin{ack}
 Thank you to Stella Biderman, Jeff Boudier, Liv Erickson, Nathan Lambert, Philipp Schmid, and anonymous reviewers for their thoughtful feedback on earlier versions of this paper. Any errors remain the first author's responsibility.

\end{ack}

\bibliographystyle{ACM-Reference-Format}
\bibliography{bey_rel}

\appendix
\section{Resourcing Calculations}
\label{Resourcing Calculations}
Calculations for local and self-hosting open-weight models depend on hardware, server, and energy costs.  For local and self-hosting, the general formula for annual cost is (hardware + (GPU energy demand in kW x cost of energy in kWh x 8760h) + maintenance/operation)) x user demand or utilization. We examine costs for local hosting. When purchased from NVIDIA, each H100 GPU costs approximately \$25,000. Server costs range widely. Self-hosting costs depend on the hardware selected, which is often priced per GPU hour. Energy demands for H100 GPUs can reach 700W, and energy costs vary by region. Energy costs generally depend on model type and task, such as generation \cite{luccioni_power_2024}. Average nationwide energy costs in the U.S. are \$0.177 per kWh, but go up to \$0.381 per kWh in San Francisco as of August 2024 \cite{noauthor_average_2024}. 

\textbf{Llama 3.1 405B Instruct}
The required number of GPUs and subsequent cost will increase with more users. Electricity costs at 700W per GPU for 8 GPUs (in FP8) continuously running over a year will cost \$8,682.91 on average in the U.S. or \$18,690.34 in San Francisco, and for 12 GPUs (in FP16), \$13,024.39 in the U.S. and \$28,035.50. Additional maintenance costs and cost for human labor must also be factored in.

For self-hosting with the lowest cost offering on Hugging Face, each GPU is priced at \$8.25 per hour, with Llama 3.1 Instruct 405B average response time at 5 seconds for 500 input tokens at FP8 being \$0.0917 cost per 500 input tokens and 100 output tokens \cite{schmid_serverless_2024}, meaning \$18.34 per million input tokens. If continually running, daily cost would be \$1,584, with an annual cost of \$578,160 in FP8.

Self-hosting or API calls via platforms have similar price offerings. API offerings via cloud or third party services range from \$5 to \$5.33 per million input tokens and \$15 to \$16 per million output tokens. Meta also hosts Llama 3.1 Instruct for U.S. users with Meta accounts (e.g. Facebook, Instagram). 



\section{Additional Figures}
\label{Additional Figures}
\begin{figure}[h!]
  \centering
  \includegraphics[width=11.5cm]{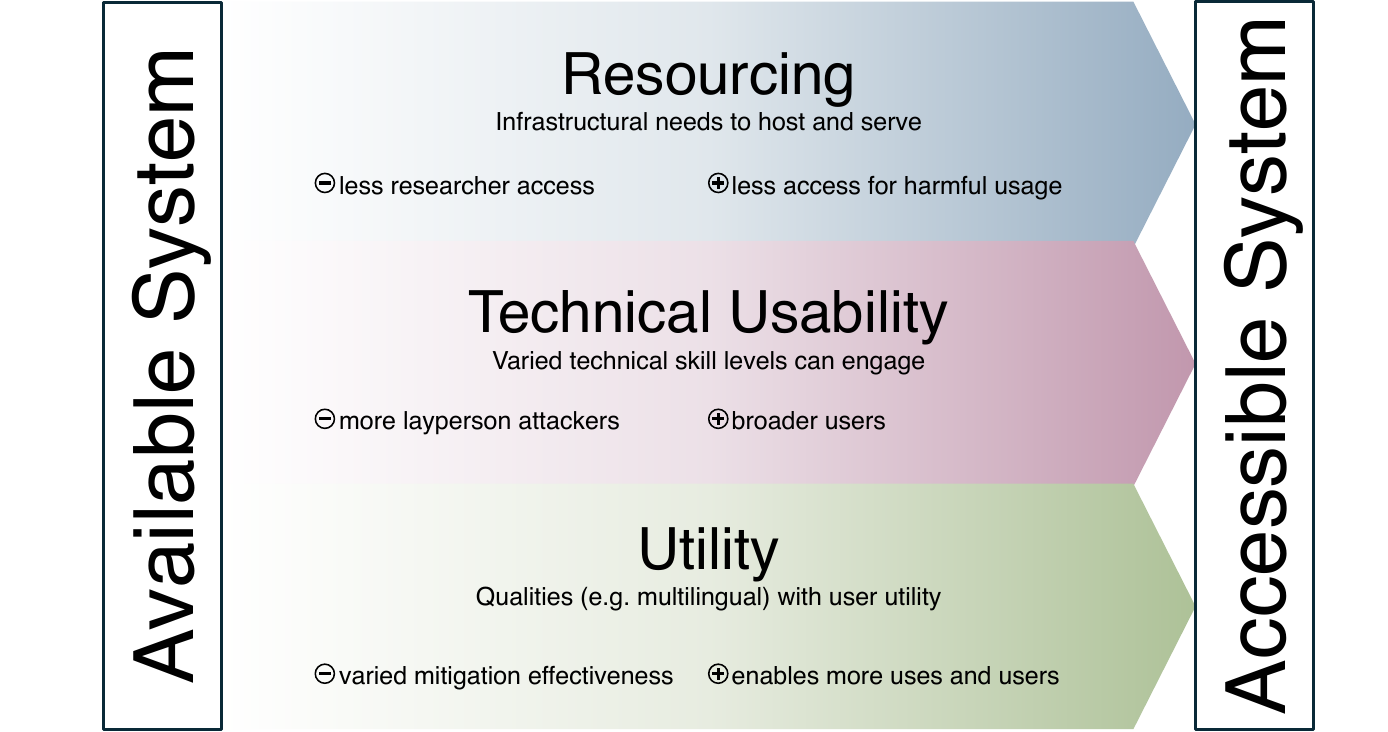}
  \caption{Shifting from Available to Accessible with High Level Tradeoffs}
\end{figure}

\end{document}